\documentclass[twocolumn,times]{aastex7}
\usepackage{CJK}

\submitjournal{ApJL}
\graphicspath{{./}{figures/}}

\usepackage{xspace}
\usepackage{amsmath}
\newcommand{\name}{2017~OF201\xspace}
\begin{document}
\begin{CJK*}{UTF8}{gkai}
\title{Discovery of a dwarf planet candidate in an extremely wide orbit: \name}

\author[0000-0002-9156-7461]{Sihao Cheng (程思浩)}
\affiliation{Institute for Advanced Study, Princeton, NJ 08540, USA}
\affiliation{Perimeter Institute for Theoretical Physics, Waterloo, N2L 2Y5, ON, Canada}
\email[show]{scheng@ias.edu}  
\correspondingauthor{Sihao Cheng}

\author[0000-0001-9592-4190]{Jiaxuan Li (李嘉轩)}
\email[show]{jiaxuanl@princeton.edu}
\affiliation{Department of Astrophysical Sciences, 4 Ivy Lane, Princeton University, Princeton, NJ 08540, USA}

\author[0009-0005-2641-1531]{Eritas Yang (杨晴)}
\affiliation{Department of Astrophysical Sciences, 4 Ivy Lane, Princeton University, Princeton, NJ 08540, USA}
\email[show]{eritas.yang@princeton.edu}

\begin{abstract}
We report the discovery of a dwarf planet candidate, \name, currently located at a distance of 90~au.
Its orbit is extremely wide and extends to the inner Oort cloud, with a semi-major axis of 830~au and a perihelion of 45~au, precisely determined from 24 observations over 20 years.
Assuming a typical albedo of 0.13, we estimate a diameter $\sim$700~km, making it the second-largest known object in this dynamical population and a dwarf planet candidate with the widest orbit.
Its high eccentricity suggests that an unseen population of similar objects would total $\sim$1\% of Earth's mass.
Notably, the longitude of perihelion of \name lies outside the clustering observed in extreme trans-Neptunian objects, posing a challenge to the proposed dynamical evidence for the hypothetical Planet Nine.
\end{abstract}

\keywords{\uat{Solar system}{1528} -- \uat{Trans-Neptunian objects}{1705} --\uat{Detached objects}{376} -- \uat{Dwarf planets}{419} -- \uat{Scattered disk objects}{1430} -- \uat{Solar system planets}{1260} }

\section{Introduction}
\label{sec:intro}

While the asteroid belt contains only less than 10$^{-3}$ of Earth's mass ($M_\oplus$), the Kuiper Belt harbors fifty times more material \citep{Pitjeva_2018}. In the vast region beyond, even greater mass is believed to reside in the scattering disk and Oort Cloud, likely totaling several $M_\oplus$ \citep{Gladman_2021}. These distant reservoirs of comets, minor planets, dwarf planets, and possibly undetected massive planets hold essential clues to the formation and evolution history of the solar system. 
Three decades ago, the first trans-Neptunian objects (TNOs) beyond Pluto was discovered \citep{Jewitt_1993}. Since then, thousands more have been identified\footnote{\url{https://www.minorplanetcenter.net/iau/lists/TNOs.html}} through various surveys, revealing a remarkable richness of this population.

However, the census of TNOs is far from complete in regions beyond about 60 au or at high ecliptic latitudes, even for large objects such as those qualifying as dwarf planets.
Due to the challenging scaling relation that the brightness of reflected light drops steeply with distance as $r^{-4}$, dedicated solar-system surveys are usually either wide (covering $\sim$10,000 deg$^2$) but shallower \citep[e.g., ][]{Trujillo_2003, Schwamb_2010}, or deeper but more focused on the ecliptic plane \citep[covering hundreds to a few thousand deg$^2$, e.g., ][]{Elliot_2005, Petit_2011, Bannister_2018, Sheppard_2016, Sheppard_2019}. 

Fortunately, recent cosmological imaging surveys, though not dedicated to solar-system science, offer a promising solution to this limitation. With wide sky coverage, deep photometry, and multiple epochs, those surveys are well-suited for detections of faint solar system objects. A prime example is the Dark Energy Survey \citep[DES;][]{DES_2016}, whose main scientific goal lies in weak lensing cosmology but has led to the discovery of more than $800$ TNOs \citep{Bernardinelli_2022}.

In this letter, we report the discovery of a large TNO, \name\footnote{\url{https://www.minorplanetcenter.net/mpec/K25/K25K47.html}}, out of a systematic search in another extra-galactic survey similar to DES but not searched before. This object is likely large enough to qualify as a dwarf planet, and its extremely wide orbit extends to the inner Oort cloud. We also discuss its migration history and interesting implications for the Planet Nine / Planet X hypothesis \citep{Batygin_2016}.

\begin{figure*}
    \centering
    \includegraphics[width=0.9\textwidth]{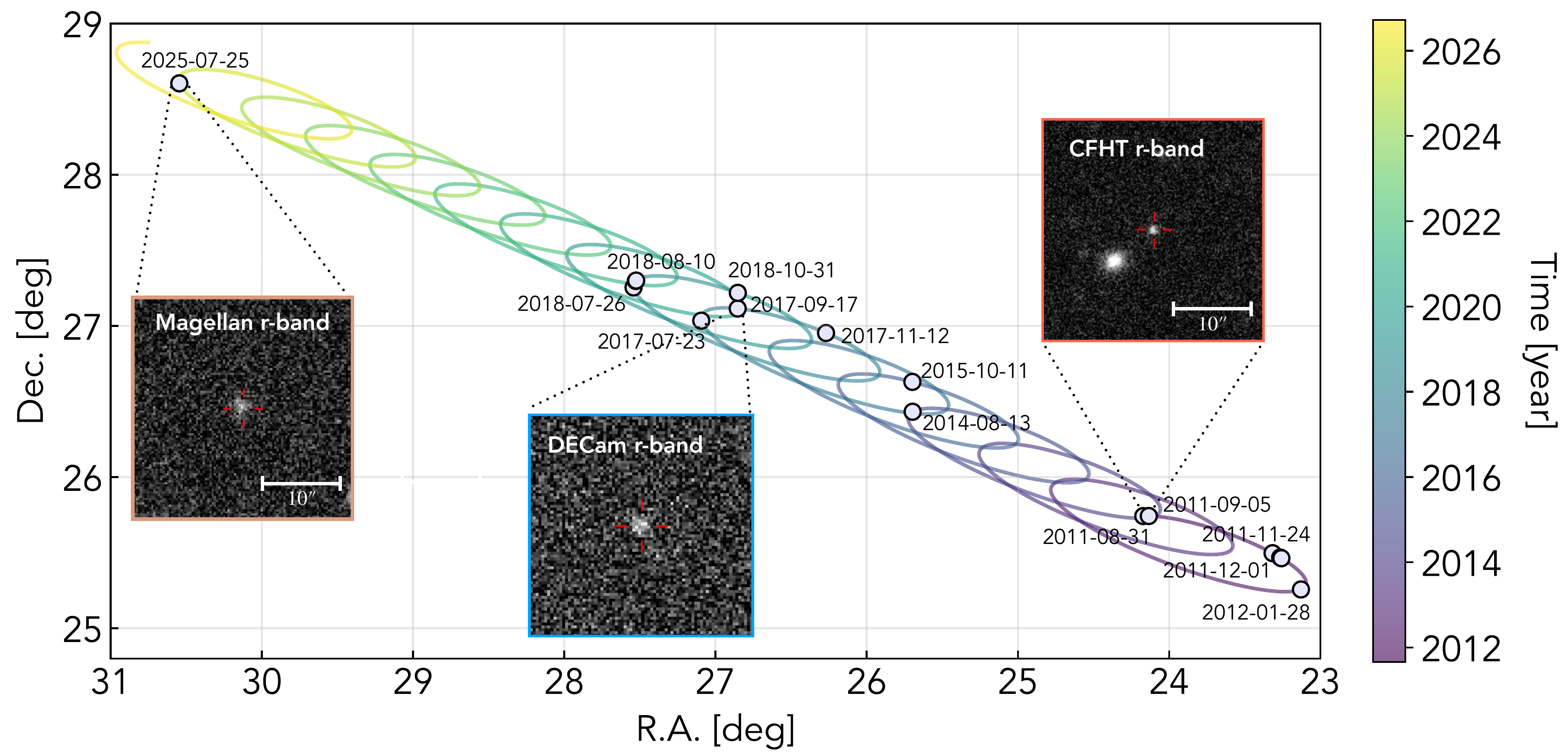}
    \caption{Trajectory of \name on the sky from 2011 to 2025, which is a combination of parallax (elliptic) and proper motion (straight) components.  Individual detections from 16 nights are shown with the predicted trajectory based on the best-fit orbit, with a small residue of 0.13 (DECam) and 0.03 (CFHT) arcsec in each component, consistent with their astrometric error. The insets show example images from Magellan, DECam, and CFHT. \\}
    \label{fig:trajectory}
\end{figure*}

\section{Discovery}
\label{sec:discovery}

We performed a systematic search for moving objects in the archival data of the Dark Energy Camera Legacy Survey \citep[DECaLS;][]{Dey_2019}\footnote{\url{https://www.legacysurvey.org/decamls/}}.
It uses the Dark Energy Camera (DECam) on the 4-meter Blanco telescope \citep{DECam}, the same facility as DES, and its sky coverage is about two times larger than DES (declination between $-15^{\circ}$ and $+32^{\circ}$). This dataset has not been searched for TNOs with its full potential, due to its sparse and irregular epoch sampling, which may separate by months to years. 

Our search algorithm is similar to \citet{Bernardinelli_2020, Bernardinelli_2022}, but optimized for the sparse epoch sampling of DECaLS. 
Details of the algorithm will be described in a future paper.  So far, \name is the most interesting object found in our search, with the largest distance and widest orbit.

After identifying 10 detections in three bands ($g,r,z$) in DECam data between 2014 and 2018, it became clear that they correspond to a single moving object with an extremely wide orbit.  With an apparent magnitude $r\sim$ 22.6~mag at 85 au away, \name has the second brightest absolute magnitude among 442 known TNOs with wide orbits ($a>$ 80 au) in the JPL Small-Body Database\footnote{\url{https://ssd.jpl.nasa.gov/tools/sbdb_query.html}}.

To further improve the orbital and photometric fitting, we used the ephemeris derived from the DECam detections to quickly search in other data archives. Following suggestions from Mike Alexandersen through a private communication, we found 9 additional $r$-band images taken in 2011 and 2012 with the 3.6 m Canada--France--Hawaii Telescope (CFHT), where \name is located exactly at the predicted positions\footnote{Taken for a TNO project by \citet{Alexandersen_2016}. It is a pity that their team missed the discovery of \name.}. The search was performed using the Solar System Object Image Search (SSOIS) tool\footnote{\url{https://www.cadc.hia.nrc.gc.ca/en/ssois/}} \citep{Gwyn_2012}.
We also searched the data archive of the Subaru and Gemini-North telescope but unfortunately found no coverage. 

After our announcement of \name, an amateur astronomer, Sam Deen, found two detections in the Sloan Digital Sky Survey \citep[SDSS;][]{York_2000}\footnote{\url{http://www.sdss.org/}}, taken as early as 2004 and 2009.\footnote{In SDSS's object catalog, these two detections have object IDs 1237666274738438899 and 1237680073395798979, misclassified as a star and a galaxy, respectively.} 
Together with our follow-up observations taken with the 6.5~m Magellan telescope on July 25, 2025, the observational arc has accumulated to 20 years with 24 observations. This unusually long baseline allows us to precisely determine its orbit right after the discovery. \\

\section{Properties of \name}
\label{sec:properties}

\begin{deluxetable}{ll}
\tablecaption{Barycentric orbital elements and photometric information of \name.}
\label{tab:parameter}
\tablehead{
\colhead{Parameters} & \colhead{Value}
}
\startdata
    $a$ (au)    &   830.1 $\pm$ 0.8      \\
    $e$         &   0.94589 $\pm$ 0.00005  \\
    $i$ (deg)   &   16.20508 $\pm$ 0.00006  \\  
    $\Omega$ (deg) &  328.5914 $\pm$ 0.0004  \\
    $\omega$ (deg) &  337.7290 $\pm$ 0.0012  \\
    $\varpi$ (deg) &  306.3204 $\pm$ 0.0013   \\
    perihelion $q$ (au)     &  44.9189 $\pm$ 0.0012   \\
    aphelion $Q$ (au)   & 1615.2 $\pm$ 1.5 \\
    period (yrs)    & 23900 $\pm$ 30 \\
    arc length & 7612 days (20.84 years) \\
    \hline
    reference epoch $t_0$ (MJD) & 60881.5 (2025-07-25)\\
    mean anomaly $M$ at $t_0$ & 1.4271 $\pm$ 0.0019\\
    distance at $t_0$ (au) & 90.7  \\
    radial velocity at $t_0$ (au/yr) & 0.64 \\
    $V$ mag at $t_0$ & 23.3 \\
    \hline
    $H_g$ & 4.10 $\pm$ 0.09 \\
    $H_r$ & 3.33 $\pm$ 0.05 \\
    $H_i$ & 2.77 $\pm$ 0.10 \\
    $H_z$ & 2.47 $\pm$ 0.10 \\
    $H_V$ & 3.72 $\pm$ 0.09 \\
    $g-r$ & 0.77 $\pm$ 0.08 \\
    $r-i$ & 0.56 $\pm$ 0.07 \\
    $B-V$ & 0.96 $\pm$ 0.09
\enddata
\end{deluxetable}

Table~\ref{tab:parameter} gives a summary of properties of \name derived from astrometric and photometric measurements of the 22 detection images (Appendix~\ref{sec:measurements}) and information read directly from SDSS's catalog.

Figure~\ref{fig:trajectory} shows the detections and best-fit orbit of \name. It is moving quickly away from us, resulting in a parallax shrink by 15\% over 20 years. This significant parallax evolution enables a precise determination of radial velocity -- the most uncertain element to measure for TNOs.  

The last time \name passed close to us was in November of 1930, and it will come back again in about 25,000 years.  Interestingly, this last perihelion was in the same year of Pluto's discovery (Feb 1930), when \name reached its maximum brightness of $V=20.2$~mag, about 4 magnitudes fainter than Pluto. Coincidentally, it was also in the same year of the establishment of the Institute for Advanced Study, where the first author of this paper is based.

\subsection{Orbit}\label{sec:orbit}

The orbit of \name is determined using the online version of the \texttt{Find\_Orb} software\footnote{\url{https://www.projectpluto.com/fo.htm}}, which implements a least-squares fitting of the orbital elements to the astrometric measurements of the 24 detections.  Thanks to the very long observation arc from 2004 to 2025, the orbit of \name is precisely determined, with the semi-major axis determined to 0.1\% precision ($a=830\pm1$~au), revealing an undoubtedly wide orbit with moderate perihelion distance ($q=44.9$~au) and low inclination ($i=16.2^\circ$).

\begin{figure*}
    \centering
    \includegraphics[width=1\linewidth]{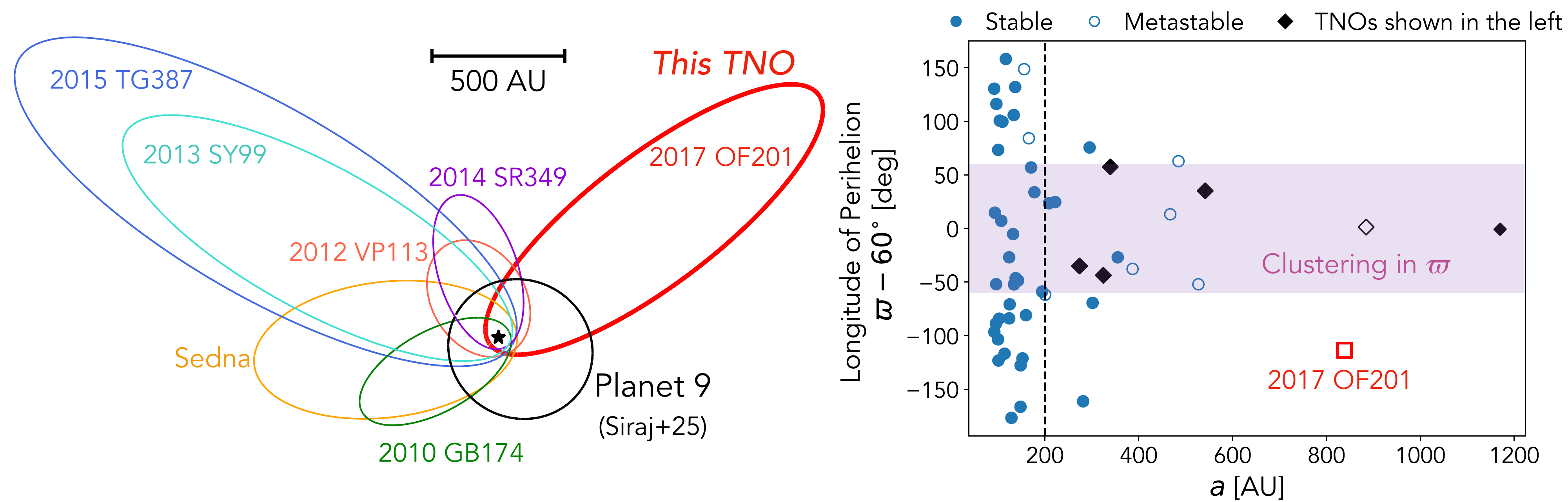}
    \caption{\textit{Left:} Plan view of the orbits of TNOs with extremely wide orbits. The newly discovered \name, highlighted in red, is an outlier to the apsidal clustering of the others. For reference, the most probable orbit of Planet 9 / Planet X from \citet{Siraj2025} is shown in black. \textit{Right:} Distribution of longitude of perihelion $\varpi$ of TNOs from \citet{Siraj2025}. The purple shaded region indicates the clustering of $\varpi$ around 60$^{\circ}$. 
    Objects with (meta)stable orbits are shown as solid (open) symbols.
    }
    \vspace{1em}
    \label{fig:orbit_view}
\end{figure*}

Figure \ref{fig:orbit_view} shows a comparison between orbits of \name and some other TNOs with extremely wide and eccentric orbits (termed ``extreme TNOs"). With such a large aphelion distance ($\sim 1600$ au), the Galactic tide may affect the orbital evolution, placing \name at the boundary between the scattering disk and inner Oort cloud.  We shall discuss its dynamical history in Section \ref{sec:4.1}.

Another notable characteristic of \name's orbit is its longitude of perihelion  $\varpi= 306^{\circ}$, an outlier to the apparent clustering around $\varpi \approx 60^{\circ}$ claimed among other extreme TNOs (Figure \ref{fig:orbit_view}, right panel).  First noticed by \citet{Trujillo_2014}, the clustering has been suggested as evidence for an undetected Planet Nine (also known as Planet X) at several hundred au \citep[e.g.,][]{Trujillo_2014,Brown_2016,Batygin_2016,Sheppard_2019, Siraj2025}. 
We shall discuss the implications of \name's orbit to Planet Nine in Section \ref{sec:P9}.

\subsection{Color and variability}

\name has a color index of $g$--$r$ = 0.77$\pm$0.08~mag, which is much redder than the Sun \citep[0.44;][]{Willmer_2018} and is consistent with other TNOs with scattering-disk and detached orbits \citep{Sheppard2010, Ofek_2012}. Though it is on the red side of the distribution, it is not yet as red as Sedna, which has $B$--$V$ = 1.24. Given the precision and time sampling of the DECam and CFHT data, we do not detect any variability over 0.1 mag, indicating that the shape of \name is likely close to spherical.  Deeper imaging with higher cadence is still needed to probe the short-time variability due to rotation or satellite transit.

\subsection{Size and mass estimate}

We estimate the size of \name based on its observed brightness, distance, and an assumed albedo.  We first obtain the $V$-band absolute magnitude $H_V=3.7$ using the photometry and best-fit orbit (Appendix~\ref{app:HG}), and then estimate the diameter $D$ through the relation $D=2\ {\rm au} \times 10^{0.2(m_\odot-H)} / \sqrt{\rho}$, where $m_\odot$ is the solar apparent magnitude \citep{Willmer_2018} and $\rho$ the geometric albedo.  As both $D$ and $\rho$ are unknown and there is only one observable $H$, ones needs an additional assumption to solve $D$. One way is the use the observed tight correlation between $D$ and $\rho$ of scattering disk objects \citep{Gerdes_2017, Muller_2020}\footnote{Also see the first figure in \url{https://www.johnstonsarchive.net/astro/tnodiam.html}. The physical origin of this relation is that larger objects have dense enough atmospheres that may freeze and form reflecting ice.}.  We find an estimate of $\rho_V$=0.13 and $D$=700 km fits well both the correlation and observations.  This estimate is further validated by comparing it with other TNOs with similar $H$, such as 2014 UZ224 \citep{Gerdes_2017}, which has an albedo of 0.13$\pm$0.04 and diameter of 630$\pm$70 km measured from thermal emission.  Our size estimate of \name indicates that \name is highly likely large enough to achieve hydrostatic equilibrium \citep[e.g.][]{Durham_1997, Tancredi_2008} and thus qualifies as a dwarf planet.\footnote{As argued by Mike Brown (\url{https://web.gps.caltech.edu/~mbrown/dps.html}), an icy body larger than 600 km in diameter is ``highly likely" a dwarf planet. For reference, the smallest icy body known to be nearly round and thus in hydrostatic equilibrium is Saturn's satellite Mimas, which has a diameter of about 400 km.}

Future observations of thermal emission from ALMA will allow for simultaneous determination of size and albedo without the need to rely on the assumed correlation \citep[e.g.,][]{Gerdes_2017}.  Occultation is another way to directly measure size \citep[e.g.,][]{Rommel_2020}, but the next possible occultation event for stars brighter than 21 mag will be in 4 years (Nov 27 2029, 19-mag star), according to the current orbital fitting.

The discovery of such a large object with high eccentricity has interesting implications for the population of objects in trans-Neptunian space.  Throughout the orbit, \name is only close enough to be detectable in DECaLS for $\sim$0.5\% of the period, which suggests that a substantial population of about 200 similar objects -- with large sizes, wide orbits, and high eccentricities -- should exist but are just too difficult to detect due to their large distance. 

The mass of \name is estimated to be 3$\times$10$^{20}$ kg = 1/20,000 $M_\oplus$, when assuming a typical density of $1.7\ \mathrm{g\, cm^{-3}}$ according to the NASA Planetary Data System\footnote{\url{https://sbn.psi.edu/pds/resource/tnocenalb.html} and reference therein.} and a diameter of 700 km. Therefore, the population of \name-like objects would add up to order of 1\% $M_\oplus$ (about the Moon's mass). For comparison, the total mass of the classical Kuiper belt is estimated to be 1--2\% $M_\oplus$ using object counting and dynamical estimates \citep[e.g.][]{Bernstein_2004, Fraser_2014, Pitjeva_2018, Petit_2023, Bernardinelli_2025}, and that of the scattering disk is 1--10\% $M_\oplus$ \citep{Gladman_2008, Volk_2008}. Therefore, discovery of large TNOs with extremely elongated orbits such as \name and Sedna \citep{Brown_2004} reveals a significant amount of mass in the outer solar system.

\begin{figure*}
    \centering
    \includegraphics[width=0.49\linewidth]{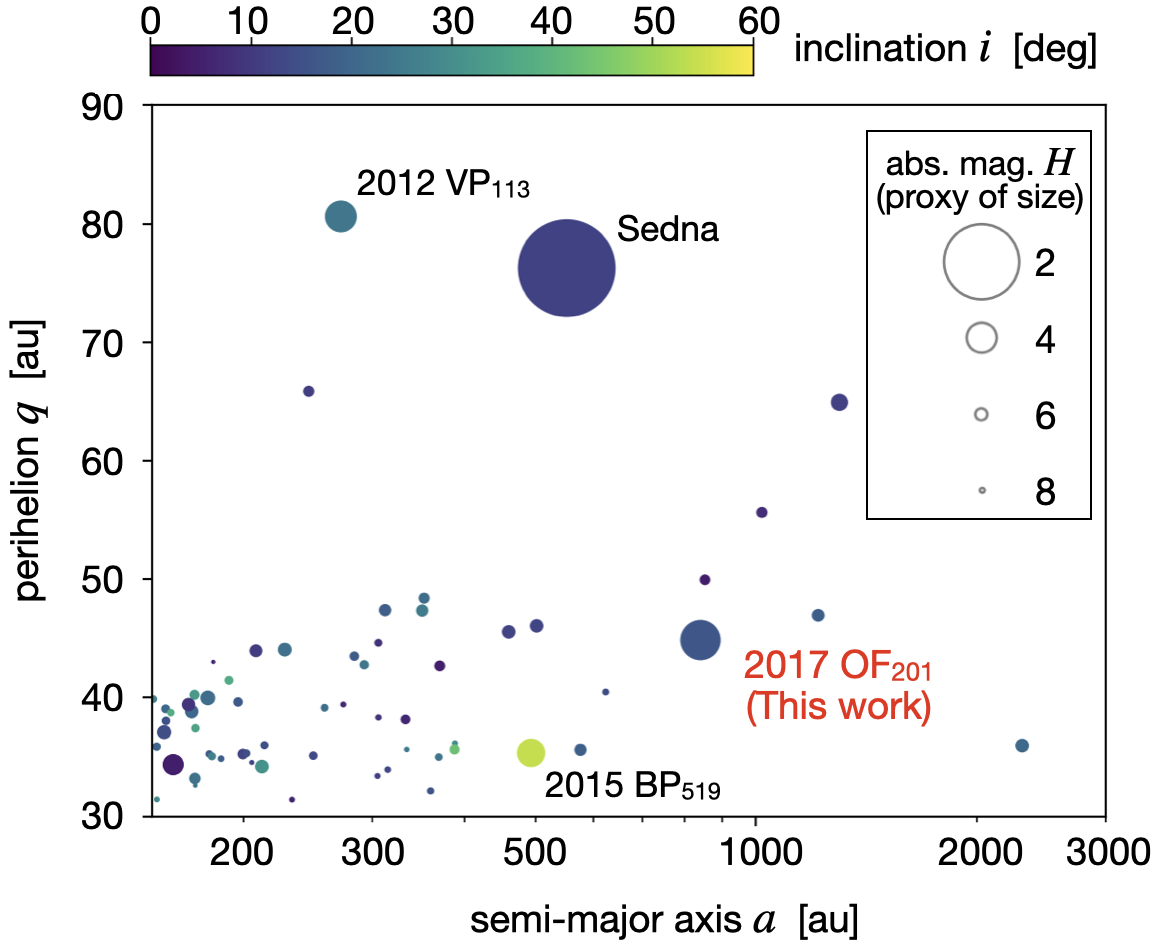}
    \includegraphics[width=0.453\linewidth]{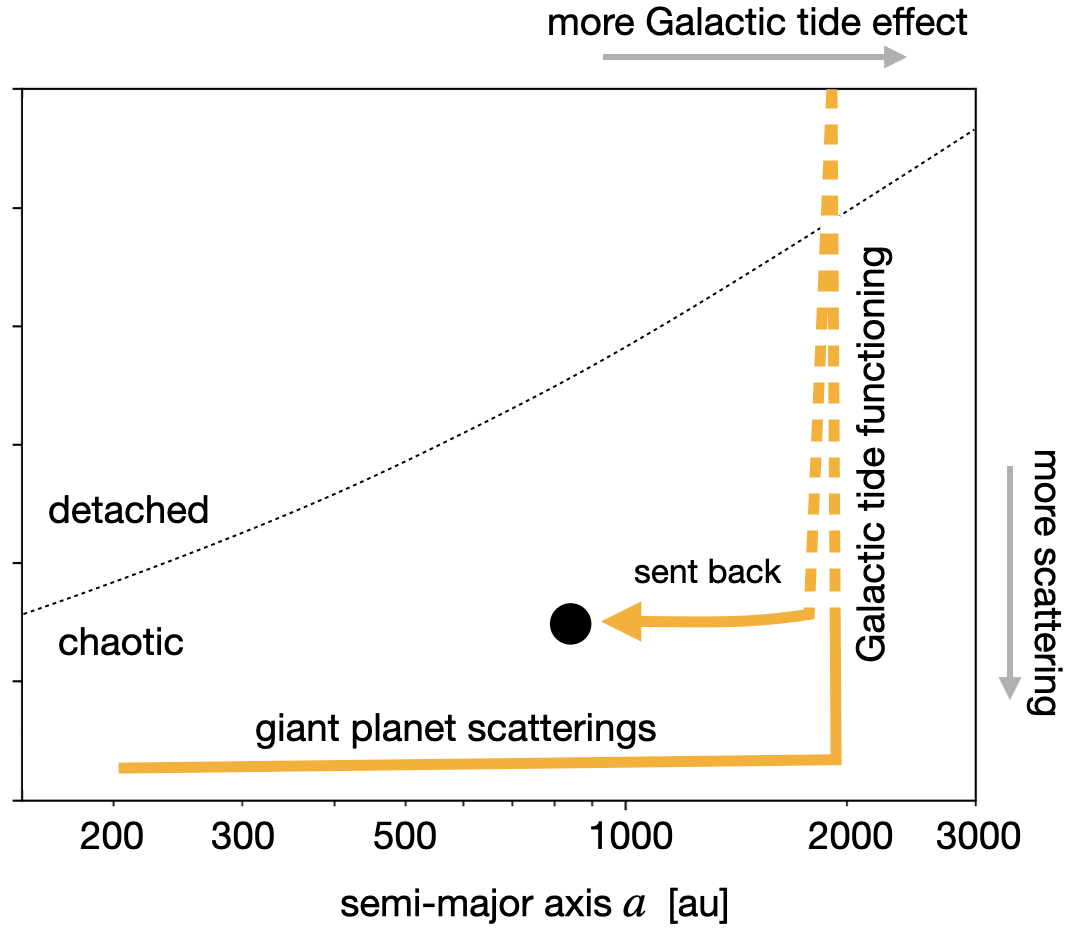}
    \caption{\emph{Left}: TNOs with wide orbits ($a>$150 au). \name is the second brightest in this population and has the widest orbit among all dwarf planet candidates. \emph{Right}: Schematic of a possible migration pathway for \name. The object may first have been scattered to a large $a$ while maintaining a low $q$ and then galactic tidal torques slowly increase $q$, detaching the orbit from Neptune. \\
    }
    \label{fig:migration}
\end{figure*}

\section{Orbital dynamics}
\label{sec:orbit_dynamics}

\subsection{Possible origin}
\label{sec:4.1}

The semi-major axis of \name evolve through chaotic diffusion driven by interactions with Neptune. However, the diffusion timescale $T$ increases steeply with perihelion distance. Following \citet{Hadden_2024}, it is approximately
\begin{equation}
T \approx 0.2~{\rm Myr} \times \exp\left[7.4\left(\frac{q}{a_{\rm N}}\right)\right] \approx 10^{10}~{\rm yr},
\end{equation}
where $a_{\rm N}$ is the semi-major axis of Neptune and $q$ is the perihelion distance of \name. This implies that \name cannot reach its current large semi-major axis through perturbations from Neptune alone, given the age of the solar system. Instead, its high-$q$, large-$a$ orbit likely reflects the influence of external torques from stellar encounters or the Galactic tide, which act during or after an initial scattering event \citep{Duncan_1987, Bannister_2017, Gladman_2021,  Batygin_2021}.

\name may have first been scattered outward by Neptune while retaining a low perihelion. Once at large $a$, the Galactic tide becomes dynamically important, gradually increasing $q$ and detaching the orbit from Neptune. As the perihelion rises into a region where planetary perturbations are weak, the object undergoes slow angular momentum oscillation and reaches a configuration like that observed today (Figure~\ref{fig:migration}). Or, alternatively, a passing star in the birth cluster of the Sun may provide the torque to raise $q$ \citep{Nesvorny_2023}.

\subsection{Long-term stability}
We investigate the long-term dynamical stability of \name using direct \textit{N}-body integrations with the \texttt{ias15} integrator \citep{ias15} in \texttt{REBOUND}\footnote{\url{https://rebound.readthedocs.io/}} \citep{Rein_2012}.
Because scattering of distant objects is primarily driven by Uranus and Neptune \citep{Batygin_2016, Gerdes_2017}, we explicitly include these planets in our simulations. Jupiter and Saturn are incorporated through a static solar quadrupole moment ($J_2$), which efficiently captures their secular influence. \footnote{For validation, we also performed computationally intensive integrations that directly included Jupiter and Saturn. The outcomes were qualitatively consistent with our simplified model.}
Given \name's large aphelion distance ($\sim1600$~au), we additionally include the vertical component of the Galactic tide \citep{Levison2001} using the \texttt{add\_force} module in \texttt{REBOUNDx}\footnote{\url{https://reboundx.readthedocs.io/}} \citep{reboundx}.

We draw 100 realizations of \name's orbits from a multivariate normal distribution based on the best-fit orbital parameters and their uncertainties, and integrate each system for 4~Gyr. At the end of the integrations, \name remains bound in more than 80\% of realizations, yet in roughly half of cases its semi-major axis varies by a factor of $\geq 2$.
Following the criterion of \citet{Batygin_2021b}, we therefore classify \name as metastable.

\begin{figure}
    \centering
    \includegraphics[width=1\columnwidth]{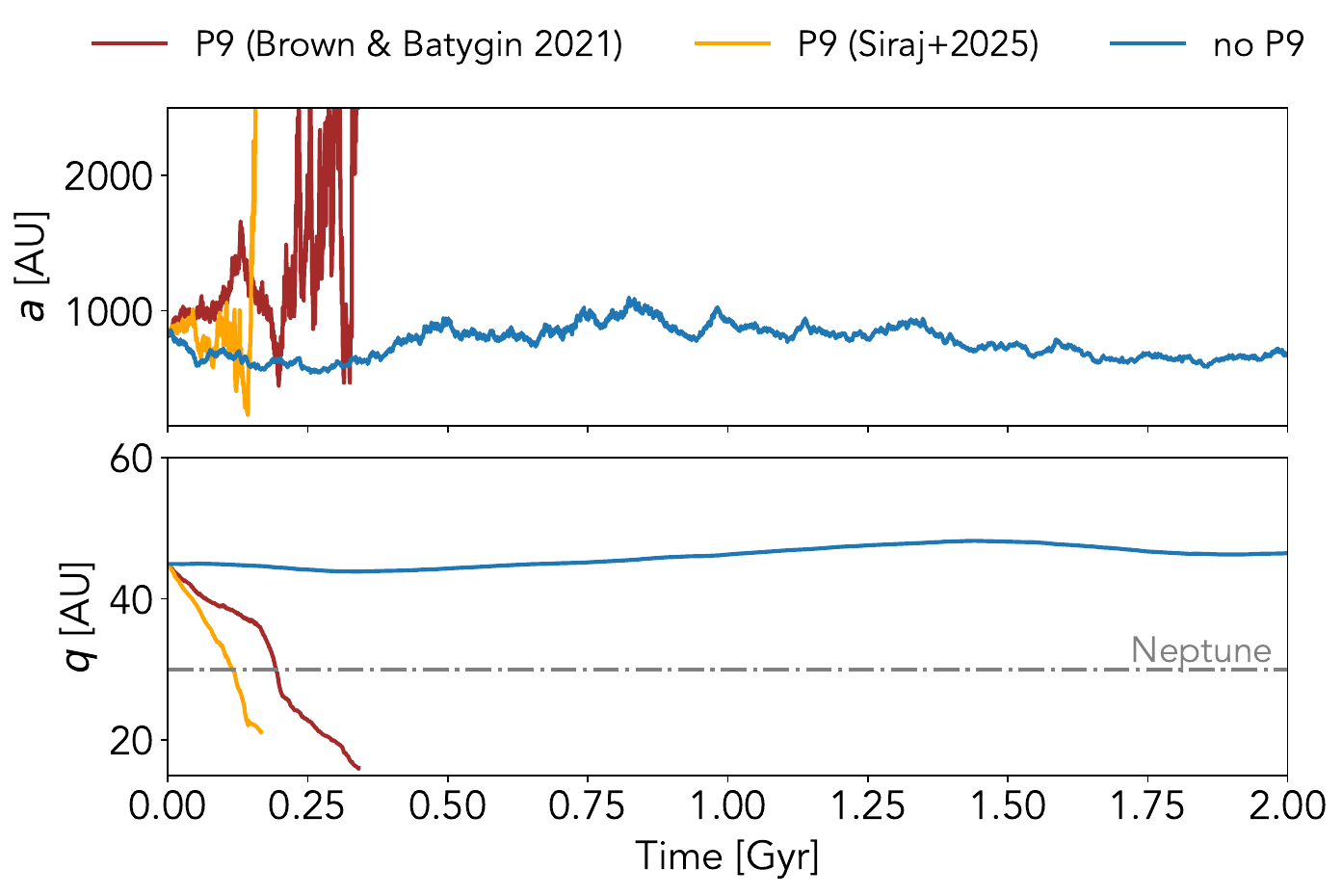}
    \caption{Typical clones of dynamical evolution simulation of \name, with and without P9. In the absence of P9, \name remains stable over 4~Gyr under the influence of the known giant planets and the Galactic tide (the x-axis is truncated for clarity). By contrast, when P9 is included, \name undergoes close encounters with Neptune and is ejected from the Solar System within $\sim$0.3~Gyr.
    }
    \label{fig:Nbody_P9}
\end{figure}

\subsection{Implications for the hypothetical Planet Nine}\label{sec:P9}
As shown in Figure~\ref{fig:orbit_view}, the longitude of perihelion of \name, $\varpi = 306^{\circ}$, lies well outside the clustering region near $\varpi \approx 60^{\circ}$ observed among other extreme TNOs. This discrepancy raises the question of whether \name is dynamically compatible with the Planet Nine (P9) hypothesis, which posits that a distant massive planet shepherds TNOs into aligned orbital configurations.

\citet{Brown_2021} inferred a candidate P9 with mass $m_p = 6.2,M_\oplus$, semimajor axis $a_p = 382.4$~au, eccentricity $e_p = 0.20$, inclination $i_p = 15.6^{\circ}$, and longitude of perihelion $\varpi_p = 246.7^{\circ}$. More recently, \citet{Siraj2025} derived an updated configuration that simultaneously reproduces the observed orbital clustering and preserves the stability of already stable TNOs, yielding $m_p = 4.1,M_\oplus$, $a_p = 296$~au, $e_p = 0.339$, $i_p = 4.27^{\circ}$, and $\varpi_p = 242^{\circ}$.

To evaluate the influence of a putative P9 on \name, we performed forward \textit{N}-body integrations including P9 with each of the parameter sets described above, and compared the results against a control case without P9. For each configuration, we fixed all orbital elements of P9 to their mean values and sampled 100 values of the mean anomaly of P9 uniformly across $[0,2\pi]$, integrating the systems for 4~Gyr. With the parameters proposed by \citet{Brown_2021}, 56\% of realizations resulted in ejection of \name, whereas the updated parameters from \citet{Siraj2025} produced a 64\% ejection fraction. The corresponding mean ejection timescales are 0.31 and 0.25~Gyr, respectively. Figure~\ref{fig:Nbody_P9} shows typical clones of the simulation result.

These results suggest that the orbit of \name is difficult to reconcile with the specific P9 configurations from \citet{Brown_2021} and \citet{Siraj2025}. Therefore, \name offers an additional challenge to the P9 hypothesis, complementing other challenges such as observational selection effects and questions regarding the statistical significance of the reported clustering \citep{Shankman_2017, Bernardinelli_2020b, Napier_2021}.

\section{Conclusion}\label{sec:conclusion}

We report the discovery of a large, exotic trans-Neptunian dwarf planet candidate, \name, from the archival data of DECaLS.  It is currently located at a distance of 90~au.  When assuming a typical albedo of 0.13, its diameter is estimated to be 700 km, large enough to qualify as a dwarf planet.

Its orbit is extremely wide and eccentric ($a=830$~au, $q=44.9$~au, $i=16.2^\circ$), influenced by both Neptune's scattering and the Galactic tide at Gyr timescale, therefore representing an overlap between the scattering disk and the inner Oort cloud.

The probability for \name to be close enough and detectable is only 0.5\% given its wide and eccentric orbit.  Therefore, there should exist a population of hundreds of \name-like objects in the outer solar system, totaling 1\% of Earth's mass, a significant fraction of the scattering disk.

The orbit of \name also poses a challenge to the Planet Nine (P9) hypothesis. Previous discoveries of extreme TNOs reveal an apparent clustering in longitude of perihelion near $\varpi \approx 60^{\circ}$, motivating the idea that an undetected planet shepherds their orbits. By contrast, \name, with $\varpi = 306^{\circ}$, lies well outside this clustering region (Figure~\ref{fig:orbit_view}) and appears dynamically inconsistent with confinement by P9. Our \textit{N}-body simulations indicate that the inclusion of such a planet leads to ejection of \name in $\sim$60\% of realizations, with a mean timescale of $\sim$0.3~Gyr (Figure~\ref{fig:Nbody_P9}). Further investigation will be necessary to assess whether this outcome robustly excludes the P9 hypothesis.

\facilities{Blanco (DECam), CFHT (MegaCam), Magellan}
\software{
    \texttt{astropy} \citep{astropy}, \texttt{matplotlib} \citep{matplotlib}, \texttt{scipy} \citep{scipy}, 
    \texttt{SExtractor} \citep{SExtractor}, \texttt{PSFEx} \citep{PSFEx}, \texttt{SCAMP} \citep{SCAMP}, \texttt{astrometry.net} \citep{Lang2010}.
}

\begin{figure*}
    \centering
    \includegraphics[width=0.9\linewidth]{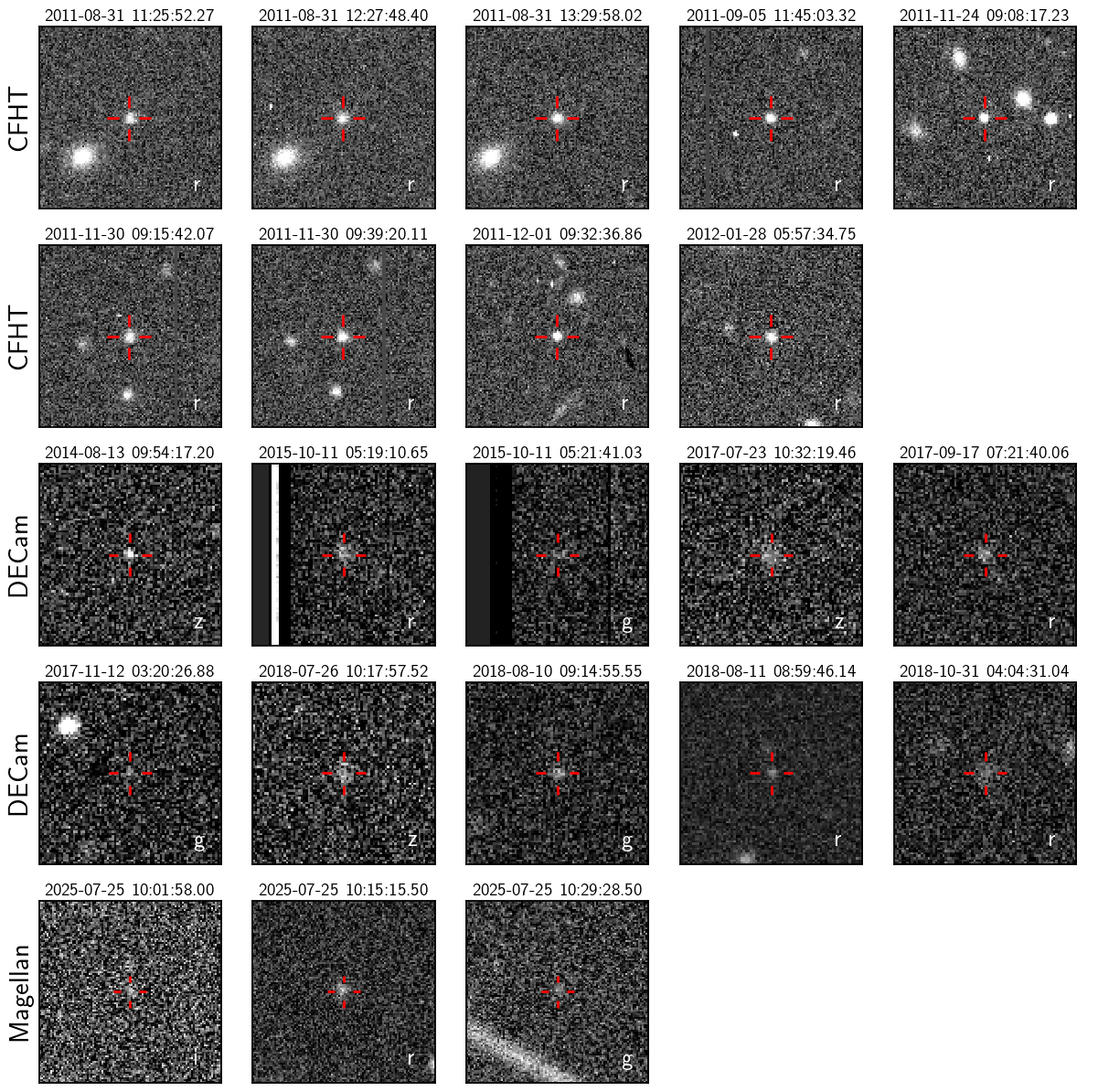}
    \caption{Cutout images of all 22 detections from CFHT, DECam, and Magellan. Each image is 20 arcsec on a side. The detection time corresponds to the middle of each exposure. CFHT data are deeper and have a much better seeing than the DECam data. The detection spans a long time baseline from Aug 2011 to Jul 2025, enabling precise determination of its orbit.}
    \label{fig:stamps}
\end{figure*}

\begin{deluxetable*}{cllllcccccc}
\tablecaption{Archival and follow-up observations of \name. The codes in parentheses are station codes of the observatory from the minor planet center. Note that the photometry reported here are measured in the native filter systems.}
\label{tab:obs}
\tablehead{
\colhead{observation time} & \colhead{R.A.} & \colhead{rms} & \colhead{Dec.} & \colhead{rms} & \colhead{magnitude} & \colhead{band} & \colhead{seeing} & \colhead{expo.} & \colhead{$\alpha$} & \colhead{$H(\alpha)$} \\
\colhead{UTC} & \colhead{deg} & \colhead{"} & \colhead{deg} & \colhead{"} & \colhead{} & \colhead{} & \colhead{"} & \colhead{s} & \colhead{deg} &
}
\startdata
\multicolumn{10}{c}{SDSS (645)} \\
\hline
2004-09-21 06:02:01.31 & 20.4237305 & 0.08   & +23.8643205  & 0.08    & 22.42 $\pm$ 0.12   &  $r$  & 1.0   & 54 & 0.40 & 3.54\\
 &  &    &   &     & 23.16 $\pm$ 0.42   &  $u$  &    &  & 4.29\\
 &  &    &   &     & 23.22 $\pm$ 0.16   &  $g$  &    &  & 4.35\\
 &  &    &   &     & 21.58 $\pm$ 0.09   &  $i$  &    &  & 2.71\\
 &  &    &   &     & 21.24 $\pm$ 0.25   &  $z$  &    &  & 2.37\\
2009-09-26 10:49:07.74 & 22.9414929 & 0.10   & +25.2131626  & 0.08    & 22.47 $\pm$ 0.12   &  $r$  & 0.9   & 54 & 0.37 & 3.43\\
 &  &    &   &     & 23.52 $\pm$ 0.84   &  $u$  &    &  & 4.48\\
 &  &    &   &     & 23.22 $\pm$ 0.17   &  $g$  &    &  & 4.18\\
 &  &    &   &     & 22.20 $\pm$ 0.14   &  $i$  &    &  & 3.16\\
 &  &    &   &     & 21.96 $\pm$ 0.41   &  $z$  &    &  & 2.92\\
\hline
\multicolumn{10}{c}{CFHT (T14)} \\
\hline 
2011-08-31 11:25:52.27 & 24.1672646 & 0.030  & +25.7422073  & 0.030   & 22.659 $\pm$ 0.038  &  $r$  & 0.9   & 320 & 0.59 & 3.543\\
2011-08-31 12:27:48.41 & 24.1669749 & 0.026  & +25.7422438  & 0.025   & 22.604 $\pm$ 0.035  &  $r$  & 0.9   & 320 & 0.59 & 3.488\\
2011-08-31 13:29:58.02 & 24.1666503 & 0.022  & +25.7422514  & 0.022   & 22.634 $\pm$ 0.032  &  $r$  & 0.8   & 320 & 0.59 & 3.518\\
2011-09-05 11:45:03.32 & 24.1300816 & 0.020  & +25.7423370  & 0.018   & 22.582 $\pm$ 0.029  &  $r$  & 0.7   & 320 & 0.55 & 3.467\\
2011-11-24 09:08:17.23 & 23.3120374 & 0.013  & +25.4947651  & 0.013   & 22.588 $\pm$ 0.022  &  $r$  & 0.5   & 320 & 0.38 & 3.471\\
2011-11-30 09:15:42.08 & 23.2618590 & 0.027  & +25.4661271  & 0.026   & 22.610 $\pm$ 0.036  &  $r$  & 0.9   & 320 & 0.44 & 3.490\\
2011-11-30 09:39:20.11 & 23.2617232 & 0.020  & +25.4660430  & 0.020   & 22.610 $\pm$ 0.030  &  $r$ & 0.7   & 320 & 0.44 & 3.490\\
2011-12-01 09:32:36.86 & 23.2539409 & 0.013  & +25.4613167  & 0.013   & 22.632 $\pm$ 0.022  &  $r$  & 0.5   & 320 & 0.45 & 3.512\\
2012-01-28 05:57:34.76 & 23.1285940 & 0.027  & +25.2561656  & 0.024   & 22.656 $\pm$ 0.038  &  $r$  & 0.8   & 384 & 0.68 & 3.508\\
\hline 
\multicolumn{10}{c}{DECam (W84)} \\
\hline
2015-10-11 05:21:41.03 & 25.694931  & 0.15   & +26.629005   & 0.18    & 23.32 $\pm$ 0.22   &  $g$  & 1.5   & 125 &  0.25 & 4.08\\
2017-11-12 03:20:26.88 & 26.267165  & 0.19   & +26.952323   & 0.22    & 24.32 $\pm$ 0.26   &  $g$  & 1.3   & 200 &  0.24 & 5.01\\
2018-08-10 09:14:55.55 & 27.525852  & 0.07   & +27.298418   & 0.08    & 23.49 $\pm$ 0.10   &  $g$  & 1.3   & 168 &  0.66 & 4.14\\
2015-10-11 05:19:10.66 & 25.694897  & 0.10   & +26.629028   & 0.10    & 22.54 $\pm$ 0.10   &  $r$  & 1.3   & 119 &  0.25 & 3.23\\
2017-09-17 07:21:40.06 & 26.850032  & 0.06   & +27.114610   & 0.07    & 22.71 $\pm$ 0.09   &  $r$  & 1.2   & 110 &  0.46 & 3.40\\
2018-08-11 08:59:46.14 & 27.523248  & 0.05   & +27.300693   & 0.05    & 22.62 $\pm$ 0.09   &  $r$  & 0.9   &  40 &  0.66 & 3.27\\
2018-10-31 04:04:31.05 & 26.850075  & 0.11   & +27.217787   & 0.12    & 22.65 $\pm$ 0.11   &  $r$  & 1.8   & 118 &  0.18 & 3.31\\
2014-08-13 09:54:17.21 & 25.694468  & 0.05   & +26.432920   & 0.05    & 21.79 $\pm$ 0.09   &  $z$  & 0.8   & 146 &  0.67 & 2.57\\
2017-07-23 10:32:19.46 & 27.094235  & 0.11   & +27.035966   & 0.09    & 21.77 $\pm$ 0.10   &  $z$  & 1.8   & 250 &  0.68 & 2.44\\
2018-07-26 10:17:57.53 & 27.541774  & 0.11   & +27.255624   & 0.10    & 22.02 $\pm$ 0.12   &  $z$  & 1.4   & 229 &  0.67 & 2.67\\
\hline
\multicolumn{9}{c}{Magellan (304)} \\
\hline
2025-07-25 10:01:58.00 & 30.547527  & 0.10   & +28.609305   & 0.10    & 22.69 $\pm$ 0.18   &  $i$  & 1.0   & 600 & 0.64 & 3.11\\
2025-07-25 10:15:15.50 & 30.547569  & 0.05   & +28.609342   & 0.05    & 23.10 $\pm$ 0.06   &  $r$  & 1.1   & 600 & 0.64 & 3.52\\
2025-07-25 10:29:28.50 & 30.547500  & 0.12   & +28.609326   & 0.12    & 23.79 $\pm$ 0.17   &  $g$  & 1.2   & 700 & 0.64 & 4.21\\
\enddata
\end{deluxetable*}

\appendix
\section{Astrometry and Photometry}
\label{sec:measurements}

We perform astrometric and photometric measurements of the object using the CFHT/MegaCam, DECam, and Magellan imaging data.  Cutouts of those images with \name centered are shown in Figure~\ref{fig:stamps}.

For the detections in CFHT, we retrieve the single-epoch images from the Canada-France-Hawaii Telescope (CFHT)/MegaCam data archive\footnote{\url{https://www.cadc-ccda.hia-iha.nrc-cnrc.gc.ca/}}. These images are all in the $r$-band with 0.5--0.9 arcsec seeing. The instrument signatures (bias, dark, and flat) have been removed in these images, but they do not yet have astrometric solutions and photometric calibrations. We implement a two-step astrometric calibration process: first obtaining an initial solution using \texttt{astrometry.net} \citep{Lang2010}, followed by fine-tuning against Gaia DR3 reference stars \citep{Gaia-dr3} using \texttt{SCAMP} \citep{SCAMP}. This procedure yields a final astrometric solution with an average positional uncertainty of $0.1\arcsec$. For photometric calibration, we first remove the sky background using \texttt{SExtractor} \citep{SExtractor} with a mesh size of $60\arcsec$. We then construct a point spread function (PSF) model using \texttt{PSFEx} \citep{PSFEx} from non-saturated point sources in each image, and perform PSF-fitting photometry using \texttt{SExtractor}. The photometric zeropoint is thus determined by comparing the instrumental magnitudes of non-saturated stars with their Pan-STARRS1 PSF magnitudes. Our object has an average $r$-band magnitude of $r=22.61$~mag in the CFHT data from 2011 to 2012. 

The DECam single-epoch images are downloaded from the Legacy Surveys\footnote{\url{https://www.legacysurvey.org/rawdata/}}, which are reduced using the DECam Community Pipeline\footnote{\url{https://noirlab.edu/science/data-services/data-reduction-software/csdc-mso-pipelines/pl206}}. These images are photometrically calibrated using Pan-STARRS1 \citep{Chambers2016} and have a preliminary astrometric solution. To ensure consistency with our CFHT analysis, we refine the astrometric calibration using \texttt{SCAMP} against the Gaia DR3 catalog. Similar to the CFHT data above, for each DECam image, we subtract the sky background with a mesh size of $60\arcsec$, generate the PSF model, and perform PSF-fitting photometry. The resulting photometry have higher uncertainties since the DECam images are shallower and have poorer seeing conditions (0.8--2.0 arcsec) than the CFHT images. 

We also include the SDSS detections on 2004 and 2009 for our orbital fitting.
In fact, those detections are listed as two objects in SDSS's catalog, with object ID 1237666274738438899 and 1237680073395798979, and misclassified as a star and a galaxy.  This is because SDSS has only single-epoch observations on most part of its coverage and therefore cannot distinguish slowly-moving objects from static ones. 
As the SDSS catalog has a high accuracy of astrometric calibration with a systematic error below 0.03\arcsec \citep{Pier_2003}, we directly use the astrometry ($r$-band) and photometry values in the catalog \citep{Fukugita_1996,Gunn_1998,Lupton_1999}.  Note that when considering photometry, one should use columns of \texttt{psfMag} instead of the default magnitude in the catalog due to the mis-classification of object type.

\section{Absolute magnitude}
\label{app:HG}

The phase angle $\alpha$ and reduced magnitude $H(\alpha)$ of each observations of \name are listed in table \ref{tab:obs}, using the best-fit orbit. The absolute magnitude $H$ is defined as the reduced magnitude $H(\alpha)=m-5\lg(r\Delta)$ when $\alpha=0$, where $\alpha$ is the phase angle, $m$ is the apparent magnitude, $r$ is the heliocentric distance, $\Delta$ is the geocentric distance. The phase curve $H(\alpha)$ reveals surface reflective properties of the object. Due to the limited range of phase angle of TNOs (0--2$^\circ$), phase curves are usually fit with a linear function $H(\alpha)=H+\beta \alpha$ \citep[e.g.][]{Rabinowitz_2007}.

For \name, only the $r$-band has good enough photometry to fit simultaneously $H$ and $\beta$, which leads to $H_r=3.33\pm0.05$ and $\beta_r=0.12\pm0.11$ per degree. The slope $\beta$ is consistent with the average value of 0.15 measured from other TNOs \citep[e.g.][]{Sheppard_2002, Rabinowitz_2007}, though the uncertainty is yet to be improved. The absolute magnitudes in other bands listed in table~\ref{tab:obs} are obtained by a least square fitting with the slope fixed to $\beta=0.12\pm0.11$ per degree.  Before fitting, all observed magnitudes from SDSS, CFHT, and Magellan \citep{McLeod_2015} have been converted into the DECam system\footnote{\url{https://des.ncsa.illinois.edu/releases/dr2/dr2-docs/dr2-transformations}}.

\begin{acknowledgments}
We thank Scott Tremaine, Yubo Su, Daniel Tamayo, Amir Siraj, Konstantin Batygin, Mike Brown, Yukun Huang, Yen-Ting Lin, Sam Hadden, Kat Volk, and Marla Geha for useful discussions and suggestions.  We thank Mike Alexandersen at the Minor Planet Center for suggesting a follow-up search in the CFHT archive. SC thanks Siyu Yao for her constant inspiration and encouragement. SC acknowledges the support of the Martin A. and Helen Chooljian Member Fund and the Fund for Natural Sciences at the Institute for Advanced Study.

This work is partly based on observations of the Legacy Surveys. The Legacy Surveys consist of three individual and complementary projects: the Dark Energy Camera Legacy Survey (DECaLS; Proposal ID \#2014B-0404; PIs: David Schlegel and Arjun Dey), the Beijing-Arizona Sky Survey (BASS; NOAO Prop. ID \#2015A-0801; PIs: Zhou Xu and Xiaohui Fan), and the Mayall z-band Legacy Survey (MzLS; Prop. ID \#2016A-0453; PI: Arjun Dey). DECaLS, BASS and MzLS together include data obtained, respectively, at the Blanco telescope, Cerro Tololo Inter-American Observatory, NSF’s NOIRLab; the Bok telescope, Steward Observatory, University of Arizona; and the Mayall telescope, Kitt Peak National Observatory, NOIRLab. Pipeline processing and analyses of the data were supported by NOIRLab and the Lawrence Berkeley National Laboratory (LBNL). The Legacy Surveys project is honored to be permitted to conduct astronomical research on Iolkam Du’ag (Kitt Peak), a mountain with particular significance to the Tohono O’odham Nation.

The Legacy Surveys imaging of the DESI footprint is supported by the Director, Office of Science, Office of High Energy Physics of the U.S. Department of Energy under Contract No. DE-AC02-05CH1123, by the National Energy Research Scientific Computing Center, a DOE Office of Science User Facility under the same contract; and by the U.S. National Science Foundation, Division of Astronomical Sciences under Contract No. AST-0950945 to NOAO.

This paper includes data gathered with the 6.5-meter Magellan Telescopes located at Las Campanas Observatory, Chile. 

This work is partly based on observations obtained with MegaPrime/MegaCam, a joint project of CFHT and CEA/DAPNIA, at the Canada-France-Hawaii Telescope (CFHT) which is operated by the National Research Council (NRC) of Canada, the Institut National des Science de l'Univers of the Centre National de la Recherche Scientifique (CNRS) of France, and the University of Hawaii. The observations at the Canada-France-Hawaii Telescope were performed with care and respect from the summit of Maunakea which is a significant cultural and historic site.

Funding for the SDSS and SDSS-II has been provided by the Alfred P. Sloan Foundation, the Participating Institutions, the National Science Foundation, the U.S. Department of Energy, the National Aeronautics and Space Administration, the Japanese Monbukagakusho, the Max Planck Society, and the Higher Education Funding Council for England. The SDSS Web Site is http://www.sdss.org/.

This research used resources of the National Energy Research Scientific Computing Center (NERSC), a Department of Energy Office of Science User Facility using NERSC award HEP-ERCAP0030970.

The authors are pleased to acknowledge that the work reported in this paper was substantially performed using the Princeton Research Computing resources at Princeton University, a consortium of groups led by the Princeton Institute for Computational Science and Engineering (PICSciE) and the Office of Information Technology's Research Computing.

\end{acknowledgments}

\bibliography{references}{}
\bibliographystyle{aasjournal}
\end{CJK*}
\end{document}